\documentclass[
    preprint,
    superscriptaddress
]{revtex4-2}
\usepackage{graphicx}
\usepackage{dcolumn}
\usepackage{amsmath}
\usepackage{bm}
\usepackage{siunitx}
\usepackage[utf8]{inputenc}
\usepackage[T1]{fontenc}
\usepackage{mathptmx}
\usepackage{fix-cm}

\begin{document}

\title{Superconducting levitation and control of a high-reflectivity micromirror}

\author{Philipp Koller}
\affiliation{These authors contributed equally to this work.}
\affiliation{University of Vienna, Faculty of Physics, Vienna Center for Quantum Science and Technology, 1090 Vienna, Austria}
\affiliation{Institute for Quantum Optics and Quantum Information (IQOQI) Vienna, Austrian Academy of Sciences, 1090 Vienna, Austria}

\author{Jannek J. Hansen}
\email{jannek.hansen@univie.ac.at}
\affiliation{These authors contributed equally to this work.}
\affiliation{Institute for Quantum Optics and Quantum Information (IQOQI) Vienna, Austrian Academy of Sciences, 1090 Vienna, Austria}
\affiliation{University of Vienna, Faculty of Physics, Vienna Center for Quantum Science and Technology, 1090 Vienna, Austria}

\author{David Walcher}
\affiliation{Institute for Quantum Optics and Quantum Information (IQOQI) Vienna, Austrian Academy of Sciences, 1090 Vienna, Austria}
\affiliation{University of Vienna, Faculty of Physics, Vienna Center for Quantum Science and Technology, 1090 Vienna, Austria}

\author{Tomáš Plichta}
\affiliation{Institute of Scientific Instruments of the Czech Academy of Sciences, v. v. i., Kralovopolska 147, Brno, 612 00, Czech Republic}

\author{Remi Claessen}
\affiliation{Institute for Quantum Optics and Quantum Information (IQOQI) Vienna, Austrian Academy of Sciences, 1090 Vienna, Austria}
\affiliation{University of Vienna, Faculty of Physics, Vienna Center for Quantum Science and Technology, 1090 Vienna, Austria}

\author{Martin Žemlička}
\affiliation{Institute for Quantum Optics and Quantum Information (IQOQI) Vienna, Austrian Academy of Sciences, 1090 Vienna, Austria}
\affiliation{University of Vienna, Faculty of Physics, Vienna Center for Quantum Science and Technology, 1090 Vienna, Austria}

\author{Philip Schmidt}
\affiliation{Institute for Quantum Optics and Quantum Information (IQOQI) Vienna, Austrian Academy of Sciences, 1090 Vienna, Austria}

\author{Rhys G. Povey}
\affiliation{Institute for Quantum Optics and Quantum Information (IQOQI) Vienna, Austrian Academy of Sciences, 1090 Vienna, Austria}

\author{Stefan Minniberger}
\affiliation{Institute for Quantum Optics and Quantum Information (IQOQI) Vienna, Austrian Academy of Sciences, 1090 Vienna, Austria}

\author{Stefan Putz}
\affiliation{Institute for Quantum Optics and Quantum Information (IQOQI) Vienna, Austrian Academy of Sciences, 1090 Vienna, Austria}

\author{Markus Aspelmeyer}
\affiliation{Institute for Quantum Optics and Quantum Information (IQOQI) Vienna, Austrian Academy of Sciences, 1090 Vienna, Austria}
\affiliation{University of Vienna, Faculty of Physics, Vienna Center for Quantum Science and Technology, 1090 Vienna, Austria}

\author{Michael Trupke}
\email{michael.trupke@oeaw.ac.at}
\affiliation{Institute for Quantum Optics and Quantum Information (IQOQI) Vienna, Austrian Academy of Sciences, 1090 Vienna, Austria}

\begin{abstract}
We introduce a method to suspend an optical micromirror, with a total mass of \SI{30}{\micro\gram}, using superconducting magnetic levitation. The micromirror is formed on a silicon membrane, coated with a high-reflectivity dielectric stack and attached to superconducting microspheres. The object is stably levitated inside a magnetic quadrupole field at cryogenic temperatures. 
Magnetic feedback on the transverse motion is used to stabilize the position of the levitator within the trapping field, allowing to measure the axial displacement of the levitator using optical interferometry. The system reaches a sensitivity of order \SI{100}{\pico\metre/\sqrt{\hertz}} near the axial trap frequency of \SI{167}{Hz}. 
This approach enables free-standing mirrors with minimal dissipation and tunable oscillation frequencies, offering a platform for precision sensing and quantum cavity optomechanics in the microgram regime.
\end{abstract}

\maketitle
\section{Introduction}
Mechanical suspension of optical elements is critical for high-precision experiments, as clamping losses and conducted vibrations degrade both mechanical coherence and displacement sensitivity.
Over the past decades, remarkable progress has been made using ultra-low-loss silica fiber suspension in gravitational-wave detectors \cite{Matichard_2015,Aasi_2015,Kawasaki_2022}, soft clamped silicon nitride structures at cryogenic and room temperature \cite{Planz_23,Kristensen2024,cupertino_2024}, crystalline mirror coatings on mechanically suspended membranes \cite{Simon_2009}, and phononic crystal resonators \cite{Chan_2011} in cavity optomechanics. Despite these advances, direct attachment to a support structure will always lead to to unavoidable dissipation channels.
Levitated systems offer a route toward removing mechanical supports entirely \cite{guccione_2013, Jiang_2020,xiao_2019} and extremely low dissipation has been demonstrated experimentally in electrical levitation of nanoparticles \cite{Dania_2024}, diamagnetic levitation of graphite sheets \cite{Tian_2024} and magnetically levitated superconducting particles \cite{Hofer_23} .

Combining diamagnetic trapping with high-reflectivity optical coatings, i.e. a "flying mirror", provides a method to suspend low-loss optical mirrors without direct mechanical attachment. 
Such a levitated mirror can undergo harmonic oscillation with a high mechanical quality factor and could, in principle, be cooled to the quantum mechanical ground state \cite{hansen_2025}.
These objects open the door to an unexplored mass regime for quantum physics experiments, where probing the gravitational interaction between massive objects in non-classical states becomes conceivable \cite{Datta2021,Miki2024,Miki2024a,Miao2020}, bridging the gap to state-of-the-art gravity measurements with classical objects \cite{Lee_2020,Westphal2021}.

Here we demonstrate controlled magnetic levitation of a microfabricated silicon mirror with a mass of \SI{30}{\micro\gram}. The levitator consists of a silicon substrate with a distributed Bragg retroreflector (DBR) and superconducting patches providing the lifting force. The device is suspended in a superconducting anti-Helmholtz trap at cryogenic temperatures (\SI{3.2}{\kelvin}). 
We resolve the harmonic motion of the translational modes and use the information to provide feedback on the motion of the levitator. 
The levitator constitutes one mirror in an interferometer, with which we measure its position with a displacement sensitivity of \SI{0.1}{\nano\meter/\sqrt{\hertz}} around the mechanical frequency of \SI{167}{\hertz}. 
This work establishes superconducting magnetic levitation as a practical suspension mechanism for high-performance optical micro-elements and a platform for cavity optomechanics without mechanical clamping.

\section{Levitator fabrication}
\begin{figure*}[t]
    \centering
    \includegraphics[width=\textwidth]{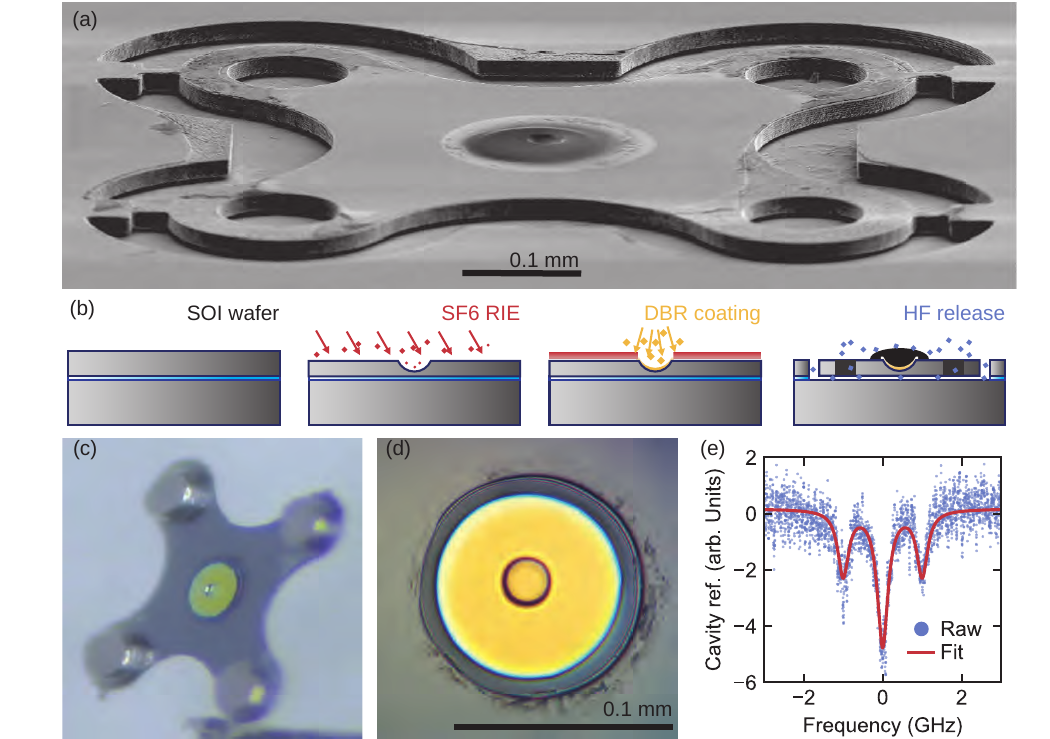}
    \caption{ Fabrication and characterization of the levitators.
    (a) Tilted SEM image of a levitator still attached to the SOI wafer, with the DBR coating in the center and four mounting holes for superconducting spheres.
    (b) Illustration of the fabrication process: The mirror substrate is prepared by etching a SOI wafer with reactive ion etch (RIE). Small deposited dielectric reflective patches (DBR Coating) are formed using a liftoff-process. The levitator shape is then etched into the device layer and the membrane is released by under-etching with hydrofluoric acid (HF). 
    (c) Completed levitator after attachment of PbSn superconducting spheres, which provide the lifting force in the magnetic trap.
    (d) Optical image of the same levitator, showing the central dielectric mirror patch (radius \SI{50}{\micro\meter}) with central wavelength of \SI{637}{\nano\meter}. In the center of the patch the parabolic profile (radius $\approx$\SI{10}{\micro\meter}) is visible. 
    (e) Sideband modulated (\SI{1}{\giga\hertz}) cavity response used for room-temperature mirror characterization of unreleased structure as in (a) by forming a high finesse ($\ge\,10^4$) Fabry–Pérot cavity.}
    \label{fig:Mircomirror_Photo}
\end{figure*}
Using established micro-fabrication techniques we manufacture high-quality mirror substrates on suspended membranes \cite{Biedermann_2010,Derntl_14,Wachter_2019,Fait_2021,Jin22}.
We start from a silicon-on-insulator (SOI) wafer with a device layer thickness of \SI{20}{\micro\meter}.
Parabolic mirror profiles, with radii of curvatures (ROC) of about \SI{120}{\micro \meter}, are etched in the device layer by a masked isotropic reactive ion etch, followed by an unmasked etch step that expands the profile defining a parabola with a depth of approximately \SI{4}{\micro\meter} and a diameter of roughly $\SI{20}{\micro\meter}$, resulting in a \SI{15}{\micro\meter} Si device layer. To polish the surface, we oxidize the substrate, by growing thermal oxide in a furnace, and subsequently remove the oxide layer by a buffer oxide etch. This procedure is performed twice. The resulting mirrors, shown in Fig. \ref{fig:Mircomirror_Photo} (c,d),  have high rotational symmetry over the central region of the parabolic mirror with $\text{ROC}_x=\SI{120\pm1}{\micro\meter}$ and $\text{ROC}_y=\SI{122\pm1}{\micro\meter}$. 
The radius of curvature of the mirrors is characterized using white-light interferometry and a two-dimensional parabolic fit. On the substrate DBR stacks with a diameter of \SI{100}{\micro\meter} are deposited, defined by a lift-off process. The coating is specified to a transmission of 10 ppm at an optical wavelength of \SI{637}{\nano\meter}. 
Restricting the coating to a localized patch reduces stress-induced deformation of the membrane substrate by the DBR coating. 

After the fabrication of the parabolic micromirrors and the deposition of the DBR coating, the levitator outline is defined by a Bosch deep reactive-ion etch of the device layer.
The levitator structure has four arms with holes to attach the  superconducting spheres and in the center is the micromirror (Fig.~\ref{fig:Mircomirror_Photo}). This shape resembles a quadcopter, offering related advantages in view of stabilization algorithms \cite{Salwa_2023}.  
The devices are released from the handle by underetching the buried oxide (BOX) layer with a buffered oxide etch over a duration of 40–60 h, while the mirror coating is protected by a resist layer. For this step, the SOI wafer is diced into individual chips. Each chip contains 36 underetched devices held by four narrow silicon tethers (\SI{\approx 10}{\micro\meter}) which can be broken to release the individual levitators from the chip. 

Finally, 90Pb10Sn ($T_c\leq\SI{7.3}{\kelvin}$) \cite{hansen_2025} superconducting spheres with a diameter of \SI{100}{\micro\meter} are press-fitted into the membrane holes, without the use of adhesives or additional bonding materials. The total mass of the assembled levitator, including silicon, dielectric coating, and four spheres, is approximately \SI{30}{\micro\gram}, dominated by the mass of the spheres with \SI{5.6}{\micro\gram} each, while the coating has a mass below \SI{0.5}{\micro\gram}. The quality of the mirror patch is verified at room temperature by forming a high finesse Fabry–Pérot microcavity between an unreleased levitator (without PbSn spheres, see Fig.~\ref{fig:Mircomirror_Photo}(a)) and a flat sapphire mirror coated with the identical dielectric stack. From the measured microcavity resonance linewidth (Fig.~\ref{fig:Mircomirror_Photo}(e)) and estimated cavity length we infer a finesse exceeding $10^4$. 

\section{Experimental setup}
\begin{figure*}[t]
    \centering    \includegraphics[width=\textwidth]{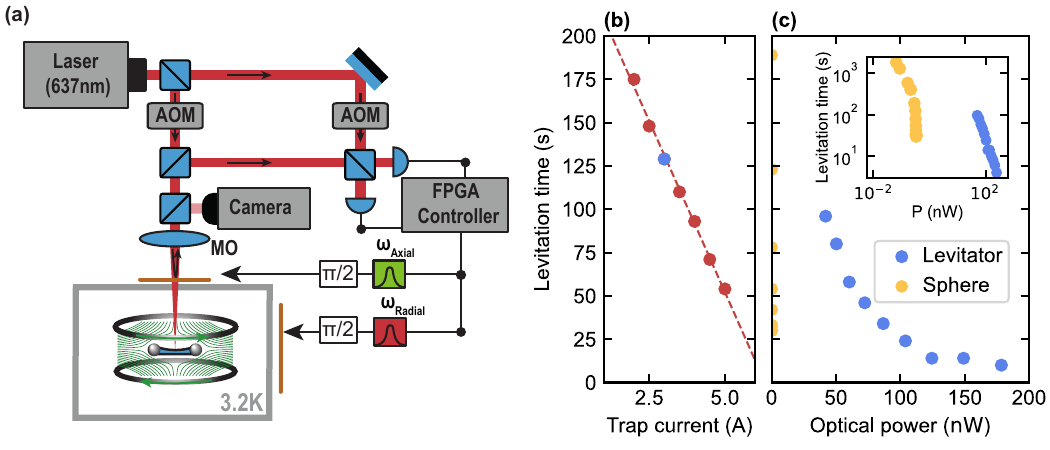}
    \caption{
    Experimental setup magnetically trapping a superconducting levitator. 
    (a) Schematic of the cryogenic and optical setup. The micromirror is levitated in a superconducting anti-Helmholtz magnetic trap inside a cryostat and probed optically through a viewport. A microscope objective (MO) focusing onto the levitator forms one arm of a Mach–Zehnder interferometer used for homodyne displacement readout. Acousto-optic modulators (AOM) in both arms enable phase locking and power control. External wires provide magnetic feedback forces for motion control. The position signal is filtered for the corresponding frequency ($\omega_{\text{Axial}},\omega_{\text{Radial}}$) and directly fed back onto the levitator with a phase delay ($\pi/2$). A camera monitors the mirror position and beam alignment.
    (b) Measured levitation time as a function of trap current with no laser light applied, showing reduced lifetime at higher magnetic fields. The dashed line is a linear fit which is a good approximation for the expected behavior in a small range of magnetic fields. 
    (c) Levitation time as a function of optical probe power at fixed trap current (\SI{3}{\ampere}), indicating additional heating from scattered and absorbed light. For comparison, the levitation time of a PbSn sphere at the same trapping current with direct illumination with an assumed detection efficiency of 10\% from reference \cite{hansen_2025} is shown. Inset: The same data on logarithmic axes.  
    }
    \label{fig:setup}
\end{figure*}

The magnetic levitation setup with optical interferometric readout is sketched in Fig.~\ref{fig:setup}(a).
The superconducting magnetic anti-Helmholtz trap is installed in a 
\textit{Montana} cryostat operated at \SI{3.2}{\kelvin}, equipped with an optical viewport from the top. The trap consists of two opposing coils separated by approximately \SI{4}{\milli\meter}, each wound with $\sim 200$ turns of superconducting copper-matrix NbTi wire with a diameter of \SI{125}{\micro\meter}. Near the trap center, the magnetic field gradients normalized per unit current are estimated to be $\bar{b}_z=\SI{24}{\tesla\per\ampere\per\meter}$ along the optical axis and $\bar{b}_{x,y}=\SI{12}{\tesla\per\ampere\per\meter}$ in the translational directions. The coils are mounted in an aluminum holder that can be opened to exchange the sample holder.
A second viewport on the \SI{30}{\kelvin} radiation shield serves as a thermal filter. It consists of a \SI{3}{\milli\meter}-thick Spectrosil window with a thermally anchored pinhole aperture of \SI{2}{\milli\meter} diameter to reduce radiative heat load from room temperature.
External copper wires outside the cryostat provide feedback control of the levitated mirror: Two off-axis coils shift the trap center in the radial plane, while a concentric coil above the viewport is used for axial feedback.
Optical access is provided by an infinity-corrected \textit{Mitutoyo} Plan Apo microscope objective (10$\times$, NA 0.28) mounted on a three-axis translation stage. The objective forms one arm of a Mach–Zehnder interferometer used for homodyne displacement readout of the back-reflected laser light with a center wavelength of \SI{637}{\nano\meter} (Toptica DL Pro). Acousto-optic modulators in both interferometer arms enable phase locking and power control in the signal arm.
A beamsplitter above the objective directs 30\% of the reflected light to a camera (DMK 37BUX287, $560\times720$ pixels) used to monitor the mirror position and beam alignment during levitation. The overall setup is described in more detail in Ref. \cite{hansen_2025}.
\section{Results}
\subsection{Levitation and lifetime}
Levitation occurs when the magnetic lifting force overcomes gravity and electrostatic adhesion, typically at trap currents of about \SI{1}{\ampere}, corresponding to around 3 times the gravitational force on the levitator. The stable equilibrium position lies in the plane perpendicular to the trap axis, coincident with the minimum of the magnetic quadrupole potential with a small offset due to gravity \cite{Hofer_2019}.

Levitation persists only while the PbSn spheres remain superconducting.
In vacuum and without any mechanical contact, heat dissipation is limited to radiation, which is negligible at cryogenic temperatures. 
Therefore, heating from room temperature radiation through the optical access results in a steady temperature increase of the levitator, leading to the quench of the superconducting spheres once the critical temperature at the applied magnetic field is reached (Fig.~\ref{fig:setup}(b,c)). This mechanism limits the maximum levitation time to $\sim$\SI{600}{\second}.
After quenching, the superconductor needs to be thermally cycled at zero magnetic field. We observe significant variation in lifetime between nominally identical devices, consistent with previously reported stochastic lift-off behavior of superconducting spheres \cite{resare2025}.

Compared to single spheres, the levitators exhibit shorter levitation times \cite{hansen_2025}, which could in part be due to the increased ratio of surface to thermal mass. 
We observe that the levitation time scales approximately linear with the aperture area of the pinhole at \SI{30}{\kelvin}, indicating that thermal radiation from room temperature is the dominant heating mechanism. Increasing trap current further reduces the levitation time, consistent with the field dependence of the critical temperature of the superconductor \cite{Chanin_1972}. 
Since the superconducting spheres are held in the magnetic field by the Si structure, the magnetic field at the mean position of each sphere is increased. 
Additionally, the deformation of the superconductor  can significantly lower the effective critical field \cite{Schmidt1997,Sato1989}.
Consequently, the levitation time depends strongly on the trap current, as illustrated in Fig.~\ref{fig:setup}(b). The maximum levitation time also scales inversely with applied optical power due to excess absorption heating and scattering shown in Fig.~\ref{fig:setup}(c).

\subsection{Mechanical modes in the magnetic trap}
The quadrupole field of the anti-Helmholtz trap confines five degrees of freedom of the levitator (three translational and two tilt), and does not constrain rotation about the symmetry axis. Nevertheless, we observe that the rotation of the levitator  is suppressed, likely due to a small asymmetry in the trapping field. The center-of-mass (c.o.m.) translational frequencies scale linearly with the magnetic field gradient and can be estimated by extending models developed for levitated superconducting spheres \cite{Hofer_2019} to the four-sphere geometry.
Due to the press-fitting process, the PbSn spheres are deformed from ideal spherical shapes. This deformation primarily increases the axial c.o.m. frequency, while the radial frequencies are slightly reduced. This behaviour is consistent with recent theoretical studies \cite{Hofer_2024,Bort_2024} and our measurements with deformed PbSn spheres (not shown). Typical devices exhibit an axial frequency increase by a factor of 1.3–1.4 compared to the spherical case.
In addition to translational motion, tilt and libration modes are observed. Tilt modes occur in a similar frequency range as axial motion and are affected by device asymmetries and individual sphere deformation.
These modes are visible in the back-reflected light when the probe beam is offset from the mirror center and also scale linearly with trap current. In the case of perfect symmetry of the trap and the levitated structure, the rotation around the optical axis is free and thereby not affected by the trap current. We observe different rotation behaviors which we believe are dependent on the initial conditions of the lift off.  

\subsection{Optical power handling of the levitator}
The sensitivity of an interferometer is ultimately dominated by laser shot noise, thus performance scales with the square root of the optical power \cite{Clerk_2010}.
Here we show improved resilience to optical probing power due to the highly reflecting optical coating compared to single spheres  themselves \cite{hansen_2025}.
To verify that the dielectric coating enables optical illumination without immediate quenching, we measure levitation lifetime as a function of applied laser power at a fixed trap current (\SI{3}{\ampere}). After active stabilization of the levitated mirror, the probe light is turned on and the time to quench is recorded.
We find that several nanowatts ($\approx \SI{3e11}{photons\per\second}$) of optical power can be applied for tens of seconds without quenching (Fig.~\ref{fig:setup}(c)). The measurement provides a lower bound on the tolerable optical power, as radial motion can displace the laser spot outside the coated region, causing rapid heating of uncoated silicon and premature quenching.

\subsection{Radial stabilization and feedback}
\begin{figure*}[t]
    \centering
    \includegraphics[width=\textwidth]{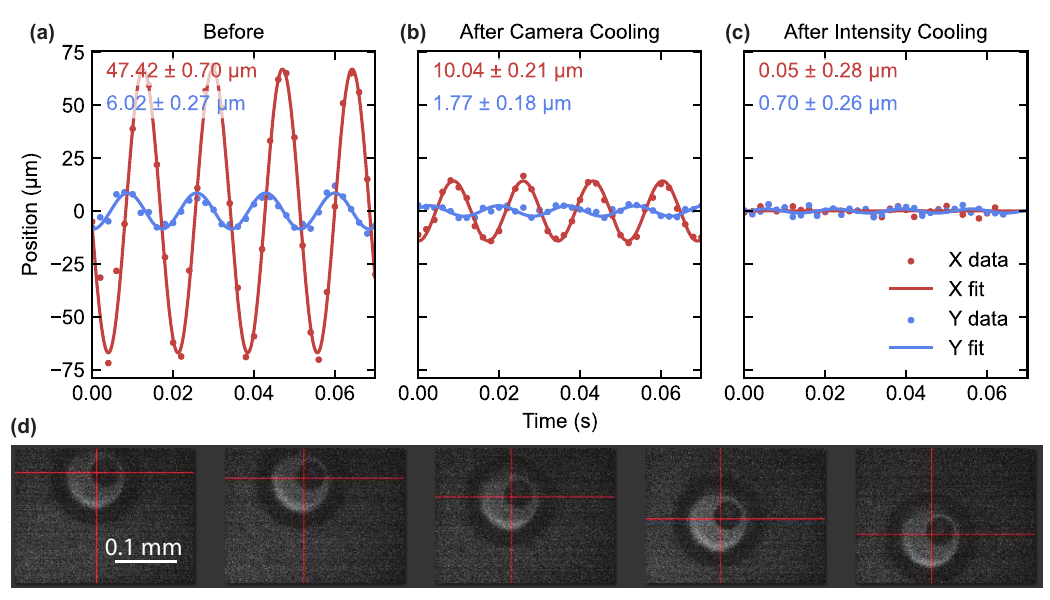}
    \caption{The radial motion of the levitator is feedback-cooled in two steps. In the first step, the center of the  mirror coating is extracted from pictures taken with a triggered camera and used to stroboscopically counteract the motion of the levitator with magnetic displacement pulses created by the external wires. In a second step the laser spot is focused into the bowl and the position-dependent intensity signal is used to control the motion of the object even more precisely using direct feedback through the same wires. (a) The motion of the levitator before cooling, extracted from the position of the bowl in the camera picture with a sinusoidal fit. (b) After the camera pre-cooling the amplitude is reduced by a factor of five, (c) the laser intensity based cooling further reduces the motion in the x-axis below the resolution of the camera based position estimation and in the y-axis to around \SI{0.7}{\micro\meter}. (d) The software based circle finder extracts the center of the mirror patch in each picture. 
    }
    \label{fig:radial_cooling}
\end{figure*}

To improve optical collection, the radial c.o.m. motion is actively stabilized using a two-stage feedback scheme adapted from \cite{hansen_2025} and demonstrated in Fig.~\ref{fig:radial_cooling}. 
We take a video of the levitator before and after each cooling step by triggering a camera at a frequency of \SI{500}{\hertz}. 
Before any cooling is done, the mirror transversely oscillates with amplitudes of \SI{47.42\pm0.7}{\micro\meter} and \SI{6.02\pm0.27}{\micro\meter} in $x$ and $y$ directions respectively in the analyzed dataset. 
Camera-based feedback cooling is performed by stroboscopically finding the center of the DBR coating with real-time image processing and applying counteracting magnetic field pulses via the feedback wires.
After feedback cooling, based on camera readout, the amplitudes are reduced to \SI{10.04\pm0.21}{\micro\meter} and \SI{1.77\pm0.18}{\micro\meter}.
This method is limited by pixel noise and mechanical frequency drifts, as it requires prior knowledge of the radial frequencies.

To overcome this detection limit we use the reflected intensity to actively cool the residual radial motion. 
As the laser beam is focused onto the concave mirror of the levitator, the reflection intensity becomes position dependent providing a linear signal for small translational offsets. We use this linear signal to perform direct cooling through a proportional magnetic force provided by the feedback wires. 
The procedure reduces the $x$ and $y$ radial displacement amplitudes to \SI{0.05\pm0.28}{\micro\meter} and \SI{0.7\pm0.26}{\micro\meter}. This method sufficiently stabilizes the optical signal reflected by the levitator such that it is suitable for interferometric detection of the axial motion.

\subsection{Axial readout and displacement sensitivity}

\begin{figure*}[t]
    \centering
    \includegraphics[width=\textwidth]{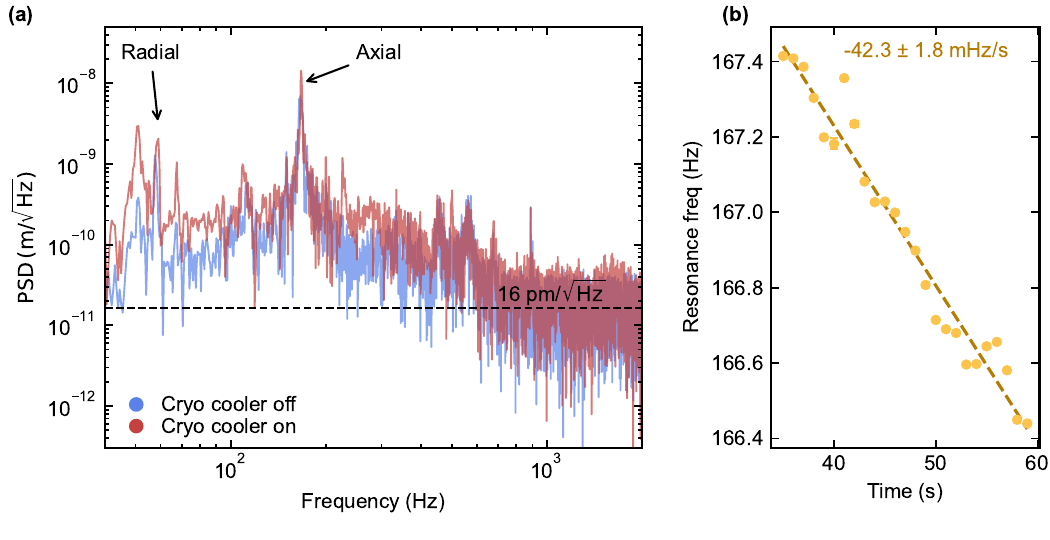} 
    \caption{We measure the power spectral density (PSD) of the interferometric signal during levitation with \SI{3.55}{\ampere} trapcurrent. (a) We resolve a prominent peak corresponding to the axial  mechanical mode at \SI{167}{\hertz}. The measurement noise floor at \SI{1}{\kilo\hertz} is around \SI{16}{\pico\meter/\sqrt{\hertz}} limited by the interferometer. The noise floor at lower frequencies is dominated by mechanical vibrations which can be reduced by switching off the cryocooler (blue curve).  (b) Mechanical resonance frequency as a function of levitation time. During levitation the mechanical frequency decreases towards lower frequency due to heating. The dashed line is a linear fit to the data.}
    \label{fig:Cryooff}
\end{figure*}

The reflected light is used for homodyne detection of the axial motion. Approximately \SI{10}{\pico\watt} of the back-reflected light is coupled into the detector. The measured displacement noise floor at \SI{1}{\kilo\hertz} is \SI{16}{\pico\meter/\sqrt{\hertz}} and around \SI{0.1}{\nano\meter/\sqrt{\hertz}} near the mechanical frequency of \SI{167}{\hertz}  as shown in Fig.~\ref{fig:Cryooff}(a).
The fundamental shot-noise limit of \SI{2.5}{\pico\meter/\sqrt{\hertz}} is not reached due to technical limitations of the interferometer and laser frequency noise. Turning off the cryocooler during levitation reduces low-frequency noise (see Fig.~\ref{fig:Cryooff}(a)) while maintaining levitation for approximately one minute, before the cryostat temperature rises above the critical temperature of the trap wires.
During levitation the axial resonance frequency, at around \SI{167}{\hertz} slowly decreases over time, as shown Fig.~\ref{fig:Cryooff}(b), with a rate of approximately \SI{42.3\pm1.8}{\milli\hertz/\second}. This behavior is consistent with gradual heating of the superconductors by thermal radiation, decreasing the restoring force on the levitator due to magnetic penetration of the superconductor. From the frequency stability and measurement resolution, we can infer only a lower bound on the mechanical quality factor of $Q\ge1500$.

\section{Discussion}
In this study we demonstrate controlled and fully superconducting magnetic levitation supporting a high reflectivity micromirror.
As for levitated spheres \cite{hansen_2025}, we can control the motion in the three translational degrees of freedom using combined techniques of camera, intensity and interferometric based readout with direct magnetic feedback. 
The smoothness and high reflectivity of the mirrors allow us to improve the measurement precision to \SI{0.1}{\nano\meter/\sqrt{\hertz}} near the motional frequency.
This microfabricated device is a step on the path to a levitated high finesse cavity which directly enhances the optomechanical coupling rate. The high reflectivity of the micromirrors allows to envisage Fabry–Pérot cavities with a finesse greater than $10^4$.
The enhancements of the intra-cavity field will allow to reduce the probe power while obtaining the same information about the mechanical state, thereby reducing the heat load on the levitator and the cryogenic environment. This approach will enhance the displacement readout sensitivity by several orders of magnitude, providing a clear path to ground-state cooling of the axial motion of the levitator \cite{Aspelmeyer_2014, Rossi_2018, hansen_2025}.

While the experiment is currently limited by extrinsic heating and vibrations of the setup, these are technical rather than fundamental constraints. They can be overcome by improved optical shielding and with vibration isolation \cite{Hofer_23,Schmidt_2024}. Replacing press-fitted spheres with microfabricated, superconducting structures \cite{Navau_2021} and incorporating chip-based, superconducting control circuits \cite{Gutierrez_2020} provides a  path toward lighter devices, improved symmetry, and integrated motion control. Operating at wavelengths greater than \SI{1}{\micro\metre} will further reduce optical absorption in silicon and allow higher optical powers.

This platform is highly versatile, as established fabrication processes allow mirror geometries to be designed as needed (flat, concave, convex, or asymmetric) \cite{Wang_2022, SAVANDER199497,Jin22,Bekker2023} and dielectric coatings to be specified for arbitrary wavelengths. Fabrication methods developed here are therefore also of interest for cryogenic spin photon interfaces \cite{Derntl_14,Awschalom_2018, kinos_2021}. The levitator platform can support any additional microfabricated structure, such as superconducting qubits or resonators, photonic \cite{Meesala_2024} and\/or phononic crystals \cite{Planz_23,Riedinger_2018}, or CMOS elements. 

To summarize, we demonstrate superconducting magnetic levitation of a low-loss dielectric micromirror formed on a concave silicon template. Compared to previous work \cite{hansen_2025}, this approach enables extended levitation timescales under laser illumination, and a significant enhancement in displacement sensitivity. These improvements mark a decisive step forward on  path to mechanical quantum states of mesoscopic objects. 

\section*{Methods}
\subsection*{Magnetic trapping model}
The levitated micromirror is confined by the quadrupole field of a superconducting anti-Helmholtz trap. The c.o.m. motion can be approximated by extending models developed for levitated superconducting spheres to a four-sphere composite object \cite{Hofer_2019}. A total levitator mass $m_{\mathrm{tot}}\approx\SI{30}{\micro\gram}$ is the sum of the 4 sphere masses ($\approx 4\times\SI{5.6}{\micro \gram}$) the silicon membrane ($\approx\SI{10}{\micro\gram}$) and the dielectric coating ($\approx\SI{0.5}{\micro\gram}$).
The superconducting volume consists of four PbSn spheres of volume $V_s$, that give the system an axial c.o.m. frequency, which can be written as
\begin{align}
f_z\approx\sqrt{\frac{3(4V_s)}{8\pi^2\mu_0 m_{\mathrm{tot}}}}\;|b_z|\; f_{\mathrm{geom}},
\end{align}
where $b_z$ is the axial magnetic field gradient, $\mu_0$ the magnetic permeability in vacuum and $f_{\mathrm{geom}}$ is a geometric correction factor which accounts for the oblateness of the pressed spheres \cite{Hofer_2024,Bort_2024}. The radial c.o.m. frequencies follow as
\begin{align}
f_{x,y}\approx\sqrt{\frac{3(4V_s)}{8\pi^2\mu_0 m_{\mathrm{tot}}}}\;|b_{x,y}|.
\end{align} 
for which the geometric factor is negligible. 
During assembly, the PbSn spheres are press-fitted into the silicon membrane and become deformed from ideal spherical shape. Previous theoretical work has already shown that such deformation primarily increases the axial trapping frequency while leaving radial motion largely unaffected. For typical devices we observe $f_{\mathrm{geom}}\approx 1.3\text{--}1.4$.

Due to the non-spherical symmetry of the composite levitator, six degrees of freedom are observed: three translational modes, two tilt modes, and one librational mode. 
Tilt modes occur in a similar frequency range as axial motion and are sensitive to asymmetries in sphere deformation and magnetic field alignment. 
Librational modes arise from residual asymmetry between the levitator and the magnetic field.

\subsection*{Superconducting lifetime and critical field}
Levitation persists only while the PbSn spheres remain in the superconducting phase. The temperature dependence of the critical magnetic field is approximated by \cite{Chanin_1972}
\begin{align}
H_C(T)\approx H_C(0)\left(1-\left(\frac{T}{T_C}\right)^2\right),
\end{align}
where $T_C$ is the critical temperature in the absence of external field. As the levitator absorbs thermal radiation, the increasing temperature reduces the critical field margin and eventually leads to quenching. 
Because the spheres are displaced from the exact magnetic field minimum by the membrane structure, the effective field experienced by the superconductor increases with trap current, leading to reduced levitation lifetime at higher gradients.

\subsection*{Lifetime measurement}
The levitation lifetime is measured by ramping the trap current to a fixed value and recording the time until the levitator drops due to a quench. 
A camera continuously monitors the mirror position to determine the drop time. 
After each measurement, the magnetic field is ramped down and the cryostat is heated above \SI{12}{\kelvin} to release trapped magnetic flux before cooling down again, ensuring reproducible starting conditions.

To evaluate heating from thermal radiation, the diameter of the radiation shield aperture at \SI{30}{\kelvin} is varied. To evaluate optical heating, the probe laser power is varied at fixed trap current (\SI{3}{\ampere}) after allowing the levitated mirror to stabilize.

\subsection*{Radial stabilization and feedback}
Radial c.o.m. motion is stabilized using a two-stage feedback scheme. First, camera images are used to determine the mirror position and apply pulsed feedback through external coils. 
In a second stage, the probe beam is focused into the concave mirror profile. 
The position-dependent intensity of the recollected light provides a linear signal for direct feedback, allowing radial motion to be reduced below \SI{1}{\micro\meter}.

\subsection*{Homodyne readout and displacement calibration\label{cavity}}
Axial motion is detected using homodyne interferometry of the back-reflected probe light. 
The displacement sensitivity is calibrated from the interferometer response and optical wavelength. The shot-noise limit is calculated from the detected optical power and compared to the measured noise floor to determine the technical noise contribution.
\subsection*{Cavity finesse characterization of micromirrors}

To verify the optical quality of the fabricated parabolic mirrors, we formed a short Fabry–Pérot cavity between an on-chip micromirror (prior to release and without PbSn spheres) and a flat sapphire mirror coated with the identical dielectric stack.
The coating consists of 34 alternating $\lambda/4$ layers of $\mathrm{Ta_2O_5}$ and $\mathrm{SiO_2}$, with a simulated transmission of approximately 10 ppm at \SI{637}{\nano\meter} on a silicon substrate and around $\approx\,$20 ppm, when deposited on sapphire substrate.

The cavity resonance was probed by scanning the cavity length with a piezo crystal while applying phase modulation sidebands at \SI{1}{\giga\hertz} to a fixed laser frequency. The sidebands served as an in situ frequency calibration to extract the resonance linewidth $\Delta \nu$, one trace is shown in Fig. \ref{fig:Mircomirror_Photo}(d).

Because the free spectral range exceeded the laser tuning range, the cavity length $L$ was estimated from calibrated microscope images of the mirror spacing. The finesse was then calculated as
\begin{align}
    \mathcal{F} = \frac{c}{2 L \Delta \nu}.
\end{align}

For larger spacings the finesse drops as the cavity mode exceeds the high-quality parabolic region of the substrate, indicating that the limitation is geometric rather than coating-related.

\section*{Funding}
This work was supported by the European Union’s Horizon 2020 research and innovation program under Grant No. 101080143 (SuperMeQ) and the European Research Council under Grant No. 951234 (ERC Synergy QXtreme). T.P. acknowledges support  from the Programme Johannes Amos Comenius under the Ministry of Education, Youth and Sports of the Czech Republic (project CZ.02.01.01/00/22\_008/0004649). We acknowledge support from the Austrian Research Promotion Agency project under projects FFG FO999914034 (SPQV) and FO999921415 (VANESSA\_QC). This work was funded by the European Union under project 101186889 (QuSPARC).
For open access purposes, the author has applied a CC BY public copyright license to any author-accepted manuscript version arising from this submission.
CzechNanoLab Project No. LM2023051 funded by MEYS CR is gratefully acknowledged  for  the  financial  support  of  the  sample  fabrication  at CEITEC Nano Research Infrastructure.
\section*{Disclosures}
The authors declare no conflicts of interest.
\section*{Acknowledgements}
The authors thank Pierre Treussart and Milan Gemaljevic for their contributions at the early stages of the project.
The authors further thank Radim Zahradníček, Peter Fecko, Thomas Astner, Gerard Higgins, Iurie Coroli and Uroš Deli\'{c} for helpful insights and inspiring discussions during the course of this work.
\section*{Data availability}
The data that support the findings of this article will be openly available. 
\
\bibliography{bib_arxiv.bib}
\end{document}